\documentclass[12pt,preprint]{aastex}
\shorttitle{Evidence of counterrotating core in the LMC}
\shortauthors{A. Subramaniam & T.P. Prabhu}
\begin{document}
\title{Evidence of a counterrotating core in the Large Magellanic Cloud}
\author{Subramaniam Annapurni \& Prabhu, Tushar P.}
\affil{Indian Institute of Astrophysics, Koramangala II Block, Bangalore - 34}
\email{purni@iiap.res.in, tpp@iiap.res.in}
\begin{abstract}
Stellar radial velocity in the central 
region of the Large Magellanic Cloud (LMC) is used to estimate the radial velocity curve along
various position angles (PA) including the line of nodes (LON). The central part of the radial velocity profile,
along the LON, shows a V-shaped profile - a clear indication of counterrotation.
The counterrotating region and the secondary bar have similar location and PA. 
The origin of the counter-rotating core could be internal(secondary bar) or external(accretion).
To explain the observed velocity profile,
we propose the existence of two disks in the inner LMC, with one counterrotating.
This two disk model is found to match the HI velocities as well. 
Two disks with different LON
and velocity profiles can create regions which are kinematically and spatially separated.
Predicted such locations are found to match the observed locations where the HI clouds are
found to have two velocities.

\end{abstract}
\section{Introduction}
The Large Magellanic Cloud (LMC) has been the subject of a large number of surveys over
the years in a wide variety of wavebands (see \citet{w90}).
The radial velocity curve of the LMC was estimated up to 8 Kpc by \citet{vahs02}
using carbon stars. 
The linear inner part of the rotation curve has one peculiarity, which is the
presence of negative velocity near the center. This result was not given any importance in
the paper due to statistically insignificant number of stars near the center.
Two kinematic components in CH stars were found by \citet{hc88} and a
lower velocity component in carbon stars was found by \citet{g00}. 
The HI velocity studies also revealed two kinematic components, the L and the D components
\citep{r84, lr92}.
Double peaked H I velocities, indicating H I clouds with two velocities in the same line
of sight have been found in some locations in the LMC, suggestive of HI gas being
located in two layers in the LMC.
The galacto-centric radial velocity curve as shown in figure 8 of \citet{r84} shows that
the central linear part of the velocity profile has a reversal of the slope near the center.
All the above, point to the possibility of a kinematically distinct component in the inner LMC. 
\citet{z03} estimated and studied
the radial velocity of 1347 stars in an attempt to detect the presence of a kinematically different
component in the inner LMC, and assigned a probability of less than 1\% for its presence.
We re-analyze the above data and search for evidence of a second kinematic
component in the inner LMC. 

The variation of stellar velocity as a function of radial distance along various position
angles (PA) is used to study the inner kinematics of the LMC. 
In a barred galaxy, the 
linear part of the curve near the center would correspond to the bar.
As we go away from the center, towards the ends of the bar, the
presence of the disk starts dominating, which is denoted by a flattening of the curve.
If a kinematically different component is present near the central region, 
one would expect to see a deviation 
from a straight line profile in the central region well inside the point of
flattening of velocity. Thus, deviations in the radial velocity curve 
near the center, estimated along various PAs, including the LON, are used to search for the 
kinematic signature of a second component within the primary bar of the LMC.
So far, no model has been proposed to explain the different kinematics shown by gas 
and stars. In this study, we propose a two-disk model for the LMC.
We model these disks with the observed inclination
and the LONs estimated from the radial velocity profiles along various PAs.
We also demonstrate that this simple
model for the LMC is capable of explaining most of the observed kinematic features.

\section{Rotation curve and the V-shaped profile} 
The radial velocity of 1347 stars presented by \citet{z03} is used 
to obtain the stellar radial velocity curves.
The bar region of the LMC is more or less covered
by this data. The main advantage of this data is that it is homogeneous such that
the same set up is used to estimate all the velocities, thereby reducing the systematic errors.
On the other hand, the observed stars do not belong to any particular evolutionary category,
thus represents a heterogeneous population. We also used
the stellar velocities of red super giants \citep{om03}, carbon stars \citep{k97} and red giants 
\citep{cole04}.
The center of the LMC is taken to be $\alpha$ = $05^h19^m38^s.0$; $\delta$ = $-69^o27'5".2$ (J2000) 
\citep{df73}, which is the optical center. The $\alpha$ and $\delta$ are converted to the 
linear X, Y co-ordinates using this center.

\citet{vahs02} estimated the LON to be 
129$^o$.9 $\pm$ 6$^o$.0, using carbon stars.
Therefore, stars located along PA = 130$^o$ and located up to 0.4$^o$ away from the PA in the
perpendicular direction, on both sides are selected. These stars were plotted as a function of
radial distance and were found to show a variation from the expected straight line profile.
This could be more clearly identified if we bin the data in the 
radial distance and study the variation of the average velocity. 
The plot is shown in figure 1, bottom left panel. 
The error bars as shown in the figure indicate the dispersion in the velocity among 
the stars in each bin. 
The striking feature of the plot is the 'V' shaped
velocity profile in the central region, where we expect a straight line profile corresponding to the
primary bar. The radial distance along any PA  is
taken positive for the northern part and negative for the southern part of the LMC \citep{f80}.
The rotation curve  as estimated from carbon stars by \citet{vahs02} (their table 2) is shown as
solid line. It can be seen that their suggestion of counterrotation near the center
is confirmed here. 

We obtained the radial velocity profile along various PAs,
to estimate the value of the PA at which maximum gradient is noticed, which will be the LON of the
central region. Therefore, we used all the
stellar data and estimated the radial velocity profiles for PAs from 0$^o$ -- 180$^o$, with a gap of
10$^o$. The slope of the counter-rotating region was found to be maximum near 120$^o$ -- 130$^o$, indicating
that this is the LON of the inner region. The stars located immediately outside the counterrotating
region, were seen not to have the LON of 130$^o$. This can be seen in the top-left
plot as shown in figure 1, where the radial velocity profile along PA = 40$^o$, which is the
minor axis for the above LON, is shown. The radial velocity has a gradient about the minor axis.
The stellar data shown here extends 
only up to 2$^o$ -- 2.5$^o$, therefore the above argument is valid up to this radial distance.
The HI velocity data of the main component from \citet{r84} 
were also analyzed in the same way and the radial velocity curve is
shown in figure 2 (lower left panel). In the case of gas, we have shown the actual data points and not the
average. The velocity pattern clearly indicates
two velocity components, where the inner region shows an s-shaped deviation. 
The PA of the LON of HI gas was found to be $\sim$ 170$^o$ \citep{lr92,k98,r84}. 
We also estimated the radial velocity curve along the HI LON
for both stars and gas and the plots are shown in the lower right panels of figures 1 and 2.
The stellar radial velocity curve shows very large scatter, which is partly due to 
less number of stars and partly due to inherent scatter in each bin. 
The HI profile, clearly shows an s-shaped deviation 
in the inner region.
NGC 3593 is also found to have similar radial velocity profile, as shown in figure 1 of \citet{b96}.
This was interpreted as a case of inner stars counterrotating with respect to the outer ones.


\section{Two disk model and the fit to the observed radial velocity profiles}
We tried to model the disk of the LMC such that
a counterrotating contribution is added to the main disk of the LMC.
The main disk of the LMC which contains the
bar and majority of the mass is responsible for most of the radial velocity curve. 
The LON of the counterrotating region is also found
to be 130$^o$. Thus the inner kinematic change is produced by invoking counterrotation,
with respect to the main large scale disk, whose parameters were estimated by \citet{vahs02}. 
On the other hand, the region just outside the counterrotating
core was found to have a LON $\neq$ 130$^o$. The position of the LON of this region can be estimated from the
fact the there is a positive slope observed at a PA of 40$^o$. This indicates that the value of LON is larger than
130$^o$. A value of 160$^o$ -- 170$^o$ for the PA was able to reproduce the observed slope along 40$^o$. 
It is found that the LON of HI gas is 170$^o$ and it extends
only up to 2.5 -- 3.0$^o$. The inclination of the HI distribution was found to be similar to that of the
stellar disk. Thus it is quite possible that the intermediate region between the
counterrotating core and the outer LMC, is dominated by a disk, which is more
of HI gas and having the above mentioned properties. The stellar data shows a lot of
scatter for 170$^o$, probably due to the fact that the stars are quite disturbed in this region.

Thus, we generated two disks, one with the LON=130$^o$ that follows the kinematics
of the stellar disk (D1) and with an inclination of $ i = 35^o$, about the LON. 
The rotational velocity for this disk is chosen such that it fits the observed range of 
stellar velocity and 
with a linear rise in the velocity up to a radial distance of 3$^o$, which stays
flat beyond that. The systemic velocity of this disk is taken as 260.0 kms$^{-1}$. 
The disk, which is dominated by HI gas, has the LON=170$^o$ (D2). 
The systemic velocity of the disk2 was taken as 275 kms$^{-1}$ \citep{k98} initially, but was found to
fit well for 260 kms$^{-1}$ as well. The rotational velocity of the disk2 is chosen such that
it fits the stellar data.
Since the two disks have the same inclination about two different LON, there will be a separation created
between the disks. As their kinematics are also different, there will be regions in the LMC, 
which have the same location in the sky, but different velocities along the line 
of sight. When one estimates the radial velocity curve, both
disks contribute such that the observed velocities could be the same as that of the individual disks or
an average of the two components, depending upon where they are located. 

To compare with the observation, we generated the radial velocity curves as shown in 
figures 1 (for stars) and 2 ( for gas).
In both the figures, the radial velocities of individual disks as well as the average are shown.
In figure 1, the profile of D1 fits the central region between $-$1.0$^o$ -- 0.5$^o$ very well
for the PA = 130$^o$. The stars in \citet{z03} mostly follow the counterrotating disk. The carbon stars
on the other hand follow the disk2 and the red giants follow the average velocity profile. Therefore,
even inside the counterrotating core, two disk model is required to fit the observed data. 
The stellar data just outside the central region, towards the north, seems
to fall closer to the average profile of D1 and D2. The assumption of the LON for the disk2 as 170$^o$
is justified from the fact that the velocity profile along the minor axis of D1 is found to fit well.
The plot at the lower right hand panel, for the PA = 170$^o$ is found to show 
a lot of scatter, which is not expected. Also, the minor axis of D2, which is PA= 80$^o$ also shows
a scatter plot. Stars located in this intermediate region, between 1$^o$ -- 2.5$^o$ 
appear disturbed.
In the case of gas, as shown in figure 2, the agreement is very good. 
For PA = 130$^o$, the HI gas is found to follow the D2 and the average. The open circles denote the
second component of HI as given by \citet{r84}. It is remarkable to see that the two disk model naturally
explains the presence of two velocities of clouds in the same line of sight.
Molecular clouds also  more or less follow
the predicted velocities. The plot for PA=40$^o$ is found to fit most of the data, except the
high velocity gas in the north and some scatter near the center. For PA = 170$^o$, which is the
LON for HI, the HI gas is found to follow D2, with some deviation near the center. 
For PA=80$^o$, molecular clouds are found to show deviation from the model, but the HI gas is
found to be more or less located within the predicted locations and does not show any significant rotation
about its minor axis. Therefore, we find that
the velocity model which was tuned to fit the stars,  fits the HI gas remarkably well.
The model velocities which were arrived are: the slope 
of D1 is 30.0 kms$^{-1}$deg$^{-1}$, up to a radius of 3$^o$, D2 has a slope of
60.0 kms$^{-1}$deg$^{-1}$, up to a radius of 0.$^o$7 and a constant velocity of 
42 kms$^{-1}$ up to 3.$^o$0. In reality, the extent of the counterrotating disk could be
much less than 3$^o$. Better data in the intermediate region is required to obtain the exact value.
A variation of up to 15\% in the above value seems to fit the
observed profile and hence the error of estimation of the slope can be considered to be 15\%.
Assuming the already estimated inclination, we find that the main disk of the LMC is
rotating in the clockwise direction. This result is similar to the finding of \citet{kb97}.
The core region has counter clockwise rotation.

In this model, there are some locations in the line of sight,
where the two disks are physically and kinematically separated.
Points in the line of sight which are located in two disks and 
separated by more than 360 pc are assumed to be
physically separated.
The scale-height of the HI disk was estimated to be about 180 pc \citep{p01}, thus the assumed
value for separation is just enough to physically separate the disks.
Locations which are in the same line of sight, but located in two disks and
have more than 20 kms$^{-1}$ difference in velocity are assumed to be kinematically separated. 
Figure 3 shows the predicted locations in the LMC, where the disks are
physically and kinematically separated. One of the observed features
in the LMC is the presence of HI clouds in layers. The location of
the double peaked clouds \citep{r84} are over plotted 
as red points. One can see that majority of the double peaked clouds are
located within the predicted region. HI in the north-western side is matched very well.
Some part of the south-east lobe of HI is found to be extended outside, and this lobe
is known to extend to larger distance from the LMC, due to tidal effects \citep{ss03}. Thus a
larger extent of the HI gas in this lobe may be expected. The presence of gas near the north-eastern
side is not explained by this model. The more or less agreeable match between the predicted and
observed locations of the double velocities indicates that the assumed two disk model 
is very close to the true nature of the LMC. 

If the microlensing in the LMC is indeed caused by self-lensing, then kinematic outliers are expected 
towards the LMC in their distribution of radial velocity, proper motion, projected distance from the
center of the LMC, distance modulus and reddening \citep{z99a,z99b}.
In the two-disk model, stars present in two disks differ in radial velocity, projected distance,
distance modulus and may be reddening also.
Though most of the stars are located in the main disk, the disk2 also hosts some
stars like the young stars.
Thus stars located in two disks can increase the star-star microlensing in the line of sight, thereby increasing
the probability of self-lensing within the inner LMC.
The locations of the observed micro-lensing events towards the LMC are shown in figure 3
as big open circles on the predicted physically and kinematically separated locations on the LMC.  
Many of the events can be found to fall within the predicted region. This suggests
that the two disks in the inner LMC fit most of the criteria required for
self-lensing within the LMC.

\section{Discussion}
The main result of the present study is in the identification of a counterrotating core
in the LMC. The inner LMC within a radius of 3$^o$ is modeled as having the presence of
two disks, with one counterrotating. 
This model is valid inside the 3$^o$ radius. The region outside is not studied here and
probably follows the parameters as estimated by \citet{vahs02}.
The two-disk model presented here is a very simple model. In reality,
the LMC may be a more complicated system, which requires detailed modeling.
The velocity data used in the present analysis was able to
bring out the counterrotation very clearly. The V-shaped profile obtained in the center should
in fact be an S-shaped profile, just like in the case of HI. The lack of data is responsible for the
V-shape, which is the truncated S-shape. 
More velocity data for stars belonging to various
population may lead to a more detailed 
model of inner and intermediate regions based on velocity dispersion, kinematics of
different populations and possibly differing rotation centers of the two disks.

Kinematically peculiar cores are generally understood as being fossil fingerprints of merging
history of host galaxies (see \citet{m98}). 
In disk galaxies it is generally interpreted as the signature of external origin of the gas component
\citep{b92}, where less than 12\% host counterrotating gas disks and less than 8\% host
counterrotating stellar disks \citep{p04}. 
The presence of counterrotation in NGC 3593 is
modeled as due to two disks, to fit the observed velocity profile \citep{b96}.
The two disk model proposed for the LMC thus closely resembles the
model for NGC 3593. 
In the LMC, the population I stars are found to show the kinematics of HI gas indicating that
they belong to the disk2. It is possible that the young stars in the LMC are
formed from the gas which has been accreted recently.
Recently, \citet{bruns04} presented
a complete HI survey of the Magellanic system. The LMC and the SMC were found to be associated with
large gaseous features - the Magellanic Bridge, the interface region and the Magellanic stream.
This gas
connects the two not only in position, but also in velocity. 
As discussed by \citet{bruns04}, a fraction of
the gas present in the vicinity is likely to be accreted by the Clouds. The gas in the Magellanic Bridge
has low velocities in the LMC-standard-of-rest frame making an accretion of some of this gas by the LMC
very likely. Thus it is very likely
that the inner gas-rich disk of the LMC could be formed from this infalling gas, which could provide
new fuel to star formation.
 
There is also a second possibility for the formation of counterrotation near the center.
The photometric location of an inner secondary bar was 
identified by \citet{as04} as $-$1.3 to 0.5 Kpc in radial distance along the LON 120$^o$. 
This is in good agreement with the extent of the counterrotating component and its PA.
Thus the identified counterrotation may be associated with the secondary bar of the LMC.
Numerical simulations of secondary bars, for example, \citet{f96} indicates that
nested, counterrotating, stable bars are viable and such systems can exist. The proposed formation
scheme is the accretion of a retrograde satellite. Counterrotating secondary bars
could also be formed due to the instabilities in the primary bar \citep{fm93}. This scenario does not
require any merger. Thus, the true nature and the reason for the formation of the counterrotating
core in the LMC is to be understood from a wide variety of possibilities which are of external/internal
origin and some discussion
in this direction can be found in \citet{corsini98}.
The proximity of
the LMC makes it ideal to test out the theories of formation of counterrotating cores and possibly understand
their origin.

We thank Andrew Cole for providing the velocity data of red giants. 

\clearpage
\begin{figure}
\plotone{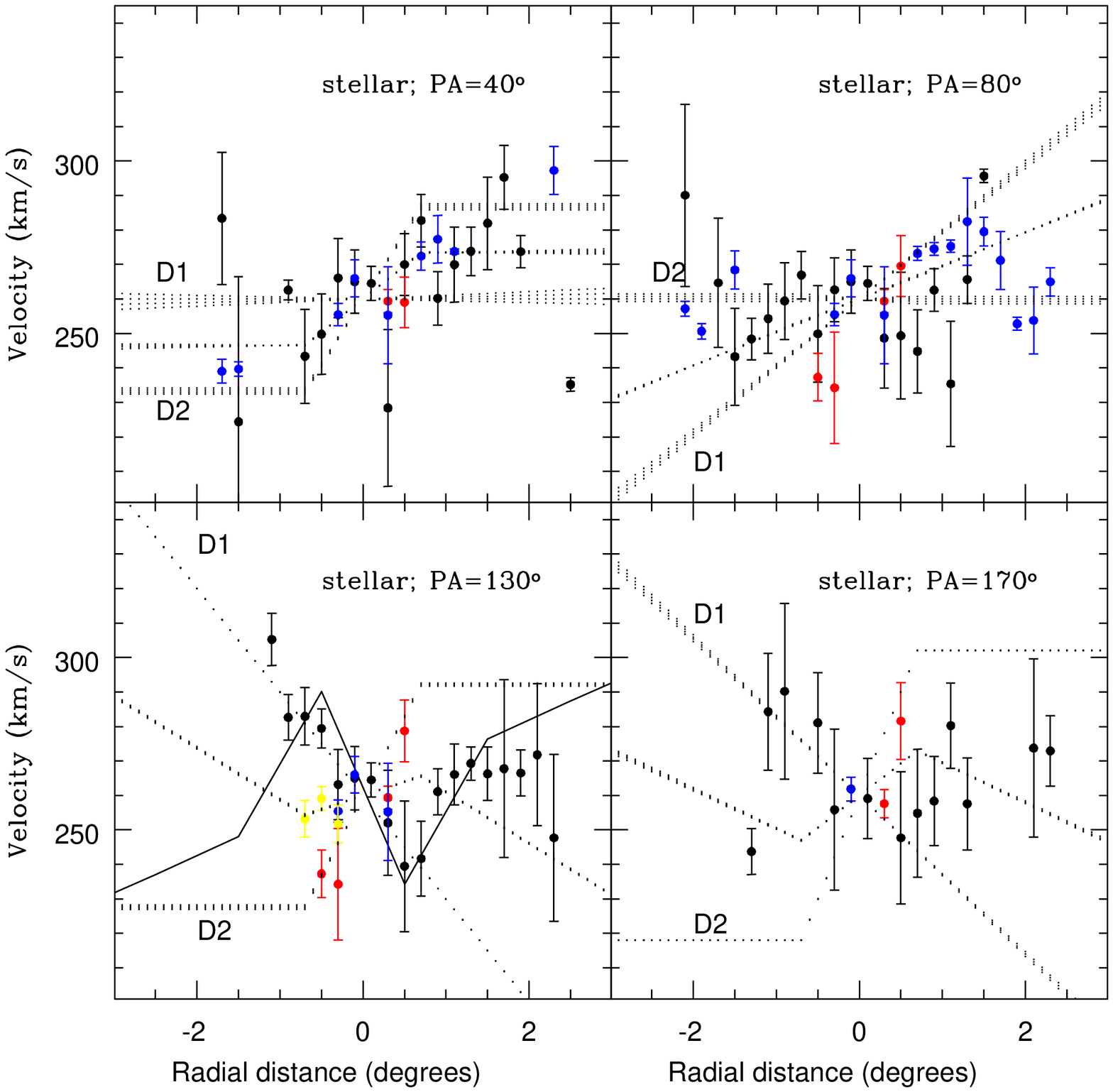}
\caption{The radial velocity profiles of stars along various PAs. Black points indicate
stellar velocities of \citet{z03}, red points indicate carbon star velocities from \citet{k97},
blue points indicate velocities of red super giants from \citet{om03} and yellow points
indicate velocities of red giants from \cite{cole04}. The bold line in the bottom left panel
is the rotation curve estimated by \citet{vahs02}. D1 and D2 denote the disk1 and disk2 respectively.
The dotted line between D1 and D2 denote the average of the two.
\label{fig1}} 
\end{figure}
\clearpage

\clearpage
\begin{figure}
\plotone{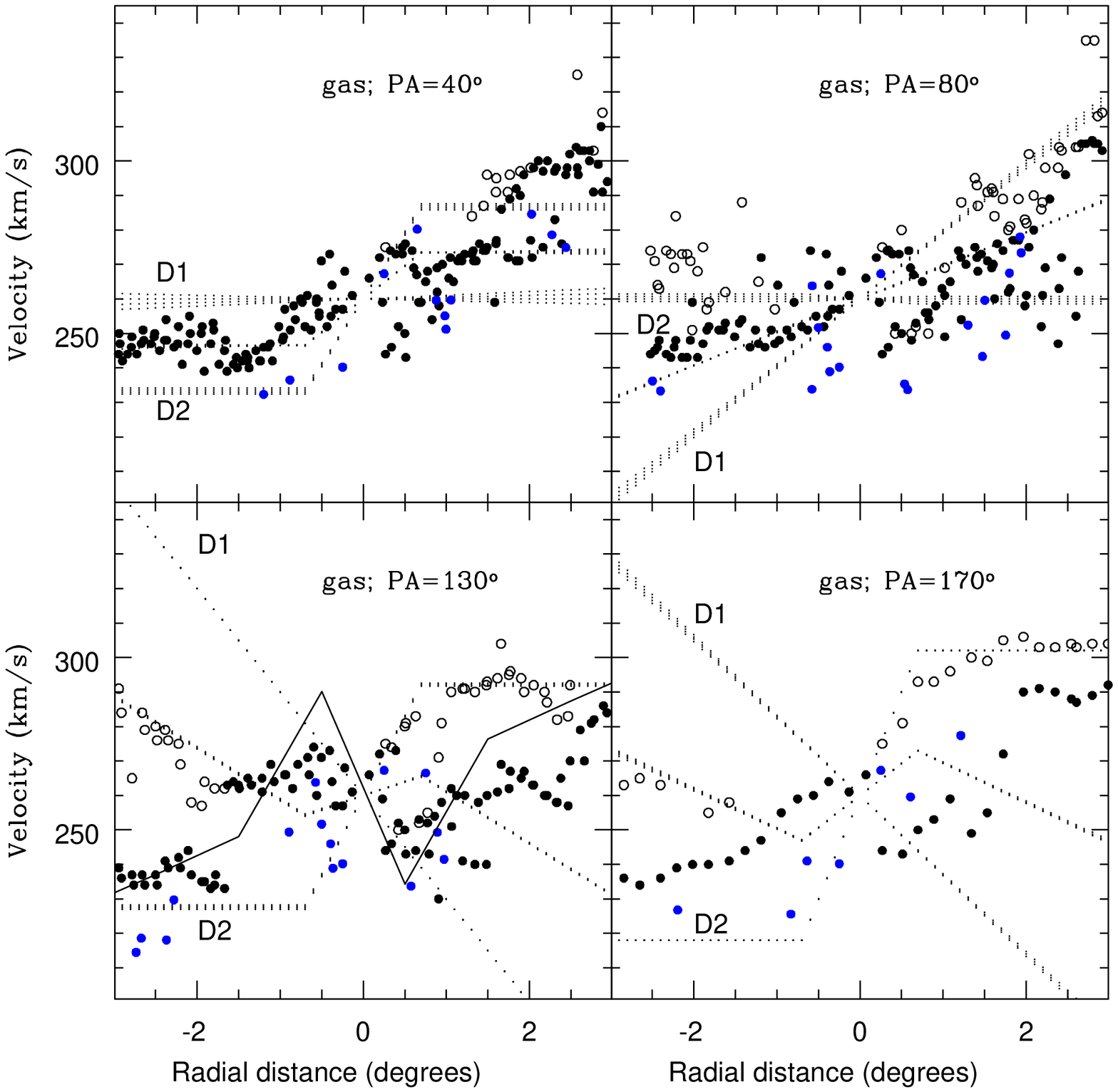}
\caption{The radial velocity profiles of gas along various PAs. Black filled points indicate
the HI data (main component) and the open circles indicate the second component of the HI gas from \citet{r84}.
The blue points indicate velocities of the molecular clouds from \citet{mizuno01}.
The bold line in the bottom left panel
is the rotation curve estimated by \citet{vahs02}. D1 and D2 denote the disk1 and disk2 respectively.
The dotted line between D1 and D2 denote the average of the two.
\label{fig2}}
\end{figure}
\clearpage


\clearpage
\begin{figure}
\plotone{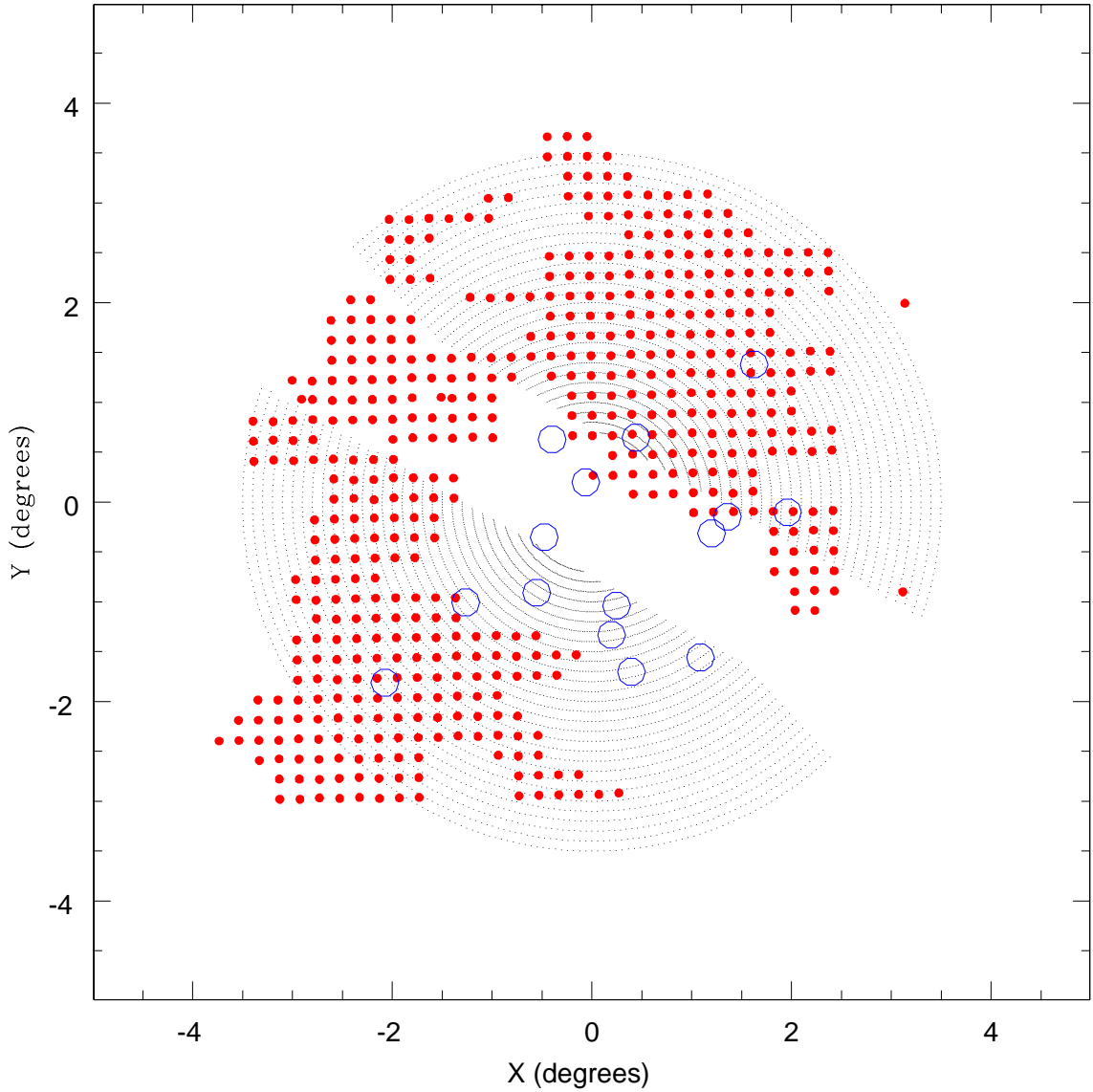}
\caption{
Locations in the LMC where the two disks are separated by more than 360 pc and have
velocity differing by more than 20 kms$^{-1}$ are shown as black dots. 
Over plotted are the locations of HI gas (red dots) where two components were detected by
\citet{r84} and the locations of the
micro-lensing events identified towards the LMC (blue circles).
\label{fig4}}
\end{figure}

\end{document}